\documentclass[twocolumn,prl,aps,epsf,showpacs]{revtex4-1} %-1

\usepackage{graphicx}
\usepackage{dcolumn}
\usepackage{bm}
\begin{document}

\title{Quantum-Classical Crossover and Apparent Metal-Insulator Transition in a Weakly Interacting 2D Fermi Liquid}

\author{Xiaoqing Zhou$^{1}$, B. Schmidt$^{1}$, C. Proust$^{2}$, G. Gervais$^{1}$, L.N. Pfeiffer$^{3}$, K.W. West$^{3}$, and S. Das Sarma$^{4}$}

\affiliation{$^{1}$ Department of Physics, McGill University,
Montreal, H3A 2T8, Canada}

\affiliation{$^{2}$ Laboratoire National des Champs Magn\'etiques Intenses, CNRS-INSA-UJF-UPS, Toulouse, 31400, France}

\affiliation{$^{3}$ Department of Electrical Engineering, Princeton University, Princeton, NJ 08544, USA}
\affiliation{$^{4}$ Condensed Matter Theory Center, Department of
Physics, University of Maryland, College Park, MD 20742, USA}

\date{\today }

\begin{abstract}

We report the observation of an apparent parallel magnetic field induced metal-insulator transition (MIT) in a high-mobility two-dimensional electron gas (2DEG) for which spin and localization physics most likely play no major role. The high-mobility metallic phase at low field is consistent with the established Fermi liquid transport theory including phonon scattering, whereas the phase at higher field shows a large insulating-like negative temperature dependence at resistances much smaller than the quantum of resistance, $h/e^2$.  We argue that this observation is a direct manifestation of a quantum-classical crossover arising predominantly from the magneto-orbital coupling between the finite width of the 2DEG and the in-plane magnetic field. 

\end{abstract}
\pacs{73.43.Qt, 75.47.De, 75.47.Gk } \maketitle

The description of a system containing many fermions in terms of a weakly interacting Fermi liquid with one-to-one correspondence with the non-interacting Fermi gas is ubiquitous in physics, applying to such diverse systems as the core of neutron stars, normal 3He and ordinary metals. The relevant temperature scale in all of these systems is set by the Fermi temperature $T_{F}$, which depends on the mass of the fermion and the density of states. Below the Fermi temperature, the system is degenerate and quantum effects due to the Pauli exclusion principle dominate.  In most metals,  $T_{F}$ is usually quite high, on the order of $10^5~K$,  so the electrical conductivity in such materials is always in the quantum degenerate regime. In semiconducting materials, where the electronic system can be constrained by design to two dimensions (2D), the Fermi temperature can be much lower due to the low density of electrons, reaching as low as tens of Kelvin. Therefore, it should  be possible, at least in principle, to observe  both the quantum degenerate ($T< T_{F}$) and classical ($T>T_{F}$) regime in a 2DEG, as long as one could neglect other spurious effects due to the presence of a lattice and the unavoidable disorder. The subject of this work is a possible experimental manifestation of a quantum-classical crossover induced in a high-mobility 2DEG by an external in-plane magnetic field, leading to an apparent 2D metal-insulator transition. In general, a quantum metal shows increasing resistivity with increasing temperature whereas a classical electron gas shows decreasing resistivity with increasing temperature, thus connecting the quantum-classical crossover with a metal-insulator transition as discussed below.
   
Applying a strong magnetic field perpendicular (along $z$) to a relatively clean 2D electronic system can have dramatic consequences on the 2D electronic transport properties, and may lead to the well-known integer \cite{Klitzing} and fractional \cite{StormerTsui} quantum Hall effects. On the other hand, applying a strong magnetic  field parallel (along $x$ or $y$) to the electron layer has much more subtle effects (that are not always well-understood). First,  the magnetic field creates an electronic spin polarization whose fraction is set by the ratio $E_z/E_{F}$, where $E_{z}=g^{*}\mu_{B} B_{\parallel}$ is the Zeeman splitting and $E_{F}$ the Fermi energy, leading to an increase in spin scattering and an increase in resistivity by no more than a factor of four \cite{DasSarma2005}. For an electronic density $n\sim 10^{11}$ $cm^{-2}$, as in the present study, this electronic spin polarization in a 2D GaAs-based system is small, less than 10\% at $10~T$. Secondly, the parallel magnetic field may also lead to a mixing of the transverse energy subbands (along z) due to the non-perturbative magneto-orbital coupling of  the finite width of the wavefunction with the field \cite{Sankar00PRL}. We have previously shown in Ref.~\cite{Zhou10PRL} that this magneto-orbital coupling can be strong in a relatively wide quantum well, and leads to a colossal magneto-resistance effect (CMR) where the resistivity increases by a factor of $\sim$300 from 0 to $45~T$.  Here, we further demonstrate that a clear change of sign of $\frac{\partial \rho_{xx}}{\partial T}$, commonly referred to as a metal-insulator transition, can be induced by a strong parallel magnetic field. We attribute this apparent MIT to a Fermi surface effect where  $T_{F}$ is suppressed by the magnetic field and the conductivity changes sign from the classical ($T>T_{F}$) to the quantum ($T<T_{F}$) regime. The MIT phenomenology here is essentially a crossover instead of a critical behaviour, and does not correspond to a typical phase transition. 

Experimentally, parallel field induced metal-insulator transition phenomenologies have been observed in a variety of relatively low density 2D systems \cite{Streda95PRB, Shahar95PRL, Pudalov97JETP, Simonian97PRL,  Yoon00PRL, Tutuc01PRL, Gao02PRL, Gunawan07NAT, Faniel, Drichko09PRB}. Similar transition phenomenologies have also been been reported in zero field when the charge carrier density was decreased below a critical value \cite{Kravchenko94PRB, Coleridge97PRB, Hanein98PRL, Simmons98PRL, Okamoto99PRL, Mills99PRL, Lilly03PRL}. In  these reports, the 2D MIT  was most often attributed to a combination of spin and localization effects, whereas the magneto-orbital coupling due to the finite thickness of the two-dimensional electron was mostly neglected, since the corresponding 2D layers were often very thin leading to negligible magneto-orbital coupling. In our work, both spin polarization and localization effects are unlikely to play a role since the  electronic spin polarization remains small in our high-mobility ultra-pure 2D system even for the highest applied magnetic fields. Likewise, the metallicity parameter $k_{F}\lambda \gg 1$ ($\lambda$ is the transport mean free path) remains large over the whole $B_{\parallel}=0-30~T$ applied parallel field range, suggesting that localization effects are unlikely. 

The unprecedented new feature in our experiment is that the parallel field driven apparent MIT occurs for a 2D system with very low disorder and very small spin polarization, in contrast to earlier observations. Our  observation of a parallel field induced metal-insulator transition has distinctive differences with those reported earlier in the literature 
\cite{Pudalov97JETP, Simonian97PRL,  Yoon00PRL, Tutuc01PRL, Gao02PRL, Gunawan07NAT, Faniel, Drichko09PRB}. First, the system we study is ultraclean and weakly interacting, and hence is well described by the Fermi liquid theory. This contrasts with previously studied low density systems in which the strong Coulomb interaction might play a crucial role \cite{Spivak10RMP}. Second, the transition occurs at a much higher temperature $\sim 10~K$ than in earlier observations ($\sim 1~K$ or below), suggesting that the contribution of phonon scattering must be much stronger in our case. Third, the resistance in the `insulating-like' phase is less than 1~$k\Omega$, and thus is much smaller than the resistance quantum $h/e^2 \approx 25$~k$\Omega$. This suggests that it is unlikely to be due to localization physics. Finally, the behavior of the measured magnetoresistance (along with the metal-insulator transition) is in qualitative agreement with a theoretical model based on the magneto-orbital coupling effect, and for which the metal-insulator transition can be well described by a field-induced Fermi surface deformation and thermal occupation. This leads us to conclude that we observe a spectacular demonstration of a quantum-classical crossover effect.  The `insulating-like' phase here is therefore not the finite temperature manifestation of the localized state in the zero temperature limit, but rather a classical system at a temperature $T$ much higher than the field-suppressed Fermi temperature $T_F$.  

The sample studied in this work is a typical high-mobility AlGaAs/GaAs/AlGaAs 2D quantum well, chosen with a charge carrier density of  $n\simeq 10^{11}$ $cm^{-2}$ and an ultra-high mobility of $\mu\simeq 10^{7}$ $cm^{2}/V\cdot s$. The corresponding $T_F\sim 40$ $K$ and the corresponding localization parameter $k_f\lambda\sim1000$ are both extremely large, showing the completely unexpected nature of  the apparent 2D MIT observed in our experiment in the presence of the applied magnetic field. The quantum well width of $40~nm$ is carefully chosen so that the magnito-orbital coupling effect can be highly non-perturbative at an accessible field of around $9~T$, in contrast to that in a $30~nm$ quantum well sample\cite{Zhou10PRL}. From Shubnikov-de Haas measurements, we have also confirmed that only the first subband is occupied at zero field, hence avoiding the complications of  second subband occupancy, which might occur in a wider quantum well. As shown in Fig.~1c, the sample has a rectangular shape with a long-to-short axis ratio $\sim$ 3:1, and for convenience the $\hat{x}$ direction is defined as parallel to the long edge. Unless otherwise stated, a measurement current of $100~nA$ was applied along a long edge, and the standard 4-wire longitudinal voltage measurement was carried out with the four  corner contacts. The measurement was confirmed to be in the linear regime ({\it i.e.}, $dV/dI$ was independent of the measurement current.)

The majority of the data were taken with a DC measurement current  in the pulsed magnetic field facility at Toulouse.  During these measurements, we were able to sweep the magnetic fields from $0~T$ to up to $60~T$ at a fixed temperature $T$. The sample was mounted on a stage that could rotate about the $\hat{y}$-axis, and the 2DEG plane of the sample was aligned parallel to the applied magnetic field. The long edge was aligned so that the condition $\textbf{\emph{I}} \parallel\textbf{\emph{B}}$ was satisfied. The temperature of the sample was controlled and monitored by using a resistive heater and a Cernox thermometer in its vicinity, using a helium bath (no more than $4.2~K$) or helium gas (above $4.2~K$) for thermal conduction. Due to the imperfection of aligning the sample, there could be a small mixing of $V_{xy}$ into $V_{xx}$, as well as other systematic errors due to the induced voltage $V_{induced}$,  notably from the eddy current arising from the pulsed field and possible thermoelectric voltage. We measured the sum of all of these contributions, {\it i.e.} $V_{total} = V_{xx} + V_{xy} + V_{induced}.$ However, by reversing the measurement current, $V_{induced}$ can be extracted and subtracted from the total signal $V_{total}$. Similarly, by reversing the magnetic field polarity, $V_{xy}$ was calibrated and minimized using the rotator. This procedure allowed us to deduce the resistance $R_{xx}=V_{xx}/I$ from $V_{total}$. Throughout this process, we estimate the overall (final) systematic error to be less than $\sim 5~\Omega$ at $\sim 20~T$ which is a few percent of the extracted resistance at this field. Finally, we have verified that the CMR effect observed at the pulsed magnet facility was in agreement (within a few percent) with the experiment in the same sample but performed at the DC magnet facility of the National High-Magnetic Field Laboratory\cite{Zhou10PRL}.

In addition to the field sweep data taken in the pulsed magnet facility, we have measured the temperature dependence of $R_{xx}$ from $4.2~K$ to $77~K$  in a superconducting magnet supplying a stable field of up to $9~T$. Good thermal conduction was maintained between the sample, the resistive heater and multiple thermometers. The $R_{xx}(T)$ at different fields were measured using a standard low frequency  (13.5~Hz) AC lock-in technique, and the measurement configuration was kept the same ({\it i.e.} $\textbf{\emph{I}} \parallel\textbf{\emph{B}}$ ). These data are found to be fully consistent with the data taken in the DC pulsed field at the same temperature. 
\begin{figure}[h]
\includegraphics[width=1\linewidth,angle=0,clip]{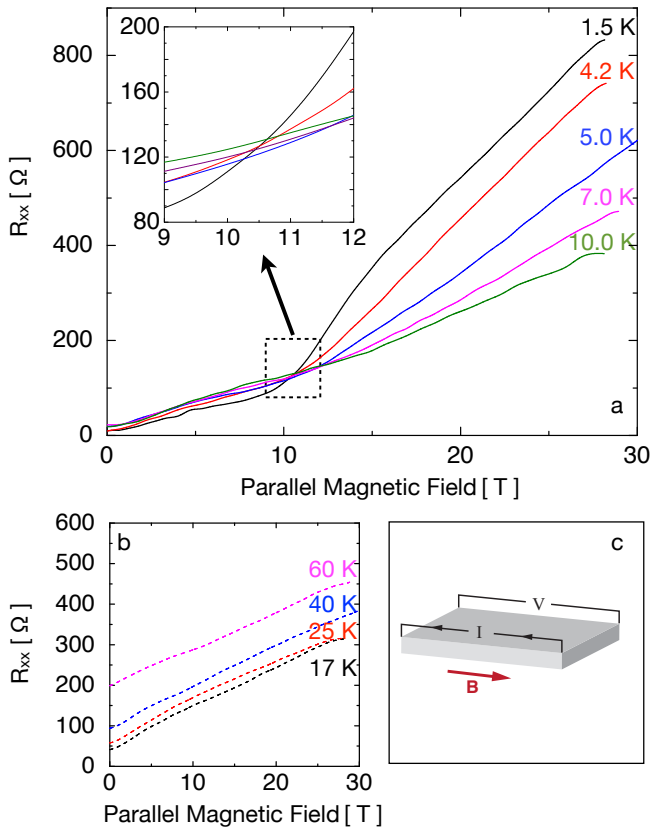}
\caption{ (a) Magnetoresistance $R_{xx}$ versus magnetic field at several temperatures from $1.5~K$ to $10~K$. The inset shows an enlarged view of $R_{xx}(B)$ in the vicinity  where different curves meet. (b) 
Magnetoresistance $R_{xx}$ at multiple temperatures from $17~K$ to  $60~K$. (c)  Sketch of the measurement configuration.}
\label{fig1}
\end{figure}
Fig.~1 shows the magnetic field dependence of $R_{xx}$ from 0 to $30~T$ at different temperatures. The lower temperature ($<10~K$) data is presented in Fig.~1a. At the lowest temperature ($\sim 1.5~K$), the magnetoresistance shows a change in slope around $11~T$ as well as $15~T$; this is fully consistent with a 2D to quasi-3D crossover transition studied in earlier work\cite{Zhou10PRL}. No magnetoresistance saturation was found for magnetic field up to 60~T (not shown). At higher temperatures, the magnetoresistance in the $\textbf{\emph{I}} \parallel\textbf{\emph{B}}$  configuration (which we focus on in this work) is enhanced at low field but reduced at high field, signaling the existence of a transition from `metallic-like' ($\frac{\partial \rho_{xx}}{\partial T}>0$) to `insulating-like' ($\frac{\partial \rho_{xx}}{\partial T}<0$) behaviour. Although no strong anisotropy is observed between the two configurations at $1~K$, this clear  transition is absent in the configuration $\textbf{\emph{I}} \perp\textbf{\emph{B}}$  in the temperature range $1-30~K$ (not shown). Finally, the fields at which different curves cross each other are found to be around $11~T$, and do not converge strictly to a single point but rather are located within a narrow window (see insert of Fig.~1a). As temperature is further increased above $\sim 10~K$ the magnetoresistance becomes more linear,  consistent with acoustic phonon scattering effects (see Fig.~1b). 

The most interesting observation is the existence of an apparent field-induced metal-insulator transition around $\sim11~T$, as shown in Fig.~2. Below $10.5~T$ the resistance increases monotonically with temperature and the slope increases above $40~K$. In fact, the temperature dependence measured at fixed magnetic fields between 0 and 9~T (see Fig.~2 inset)  suggests a temperature dependence given by the following empirical equation
\begin{equation}
R_{xx} =  a_1 + a_2T + b_1\frac{e^{-b_2/T}}{T},
\end{equation}
where the linear temperature dependence and the exponential temperature dependence are the signature of acoustic phonon scattering and optical phonon scattering, respectively \cite{sankar90PRB}.

Below $9~T$, all curves of $R_{xx}$ versus temperature are nearly parallel to each other, which implies that phonon scattering processes are not strongly affected by the parallel magnetic field (as is also indicated by the fitting parameters given in the caption of Fig.~2). Strikingly, above $10.5~T$ the temperature dependence is modified;  $\frac{\partial \rho_{xx}}{\partial T}$ changes sign and becomes negative at low temperature ($<20~K$). The appearance of this `insulating-like' behaviour ($\frac{\partial \rho_{xx}}{\partial T}<0$)  is inconsistent with equation~1. The negative temperature dependence also becomes stronger at lower temperatures. For all these data,  the metallicity parameter $k_F \lambda$ is always large since the resistance remains below $\sim 1~k\Omega$ even at the lowest temperature probed in this experiment ($\sim 1.5~K$). At higher temperatures the resistance gradually crosses over to a metallic regime ($\frac{\partial \rho_{xx}}{\partial T}>0$). In fact, at temperatures above $25 ~K$, the  temperature dependence of $R_{xx}$ mimic that with zero magnetic field. This observation suggests 
that it is dominated by the phonon scattering, seemingly largely unaffected by the applied magnetic field. 
\begin{figure}[h]
\includegraphics[width=1\linewidth,angle=0,clip]{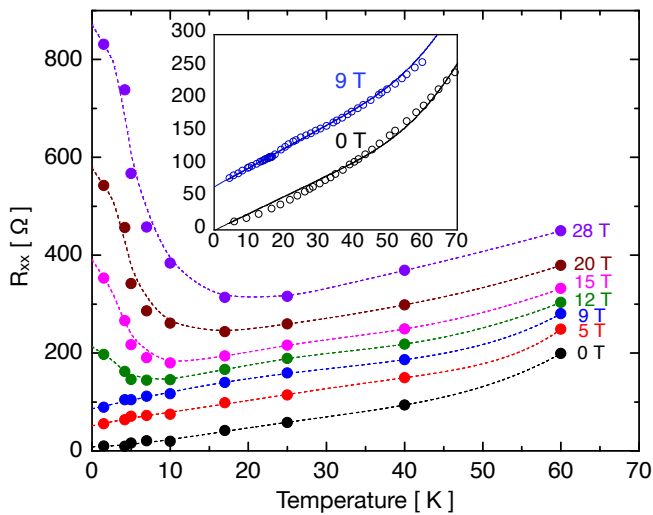}
\caption{ a) Magnetoresistance $R_{xx}$ at several fields from 0~$T$ to 28~$T$. The inset shows the temperature dependence at fixed magnetic field value of  0~$T$ (black circle) and 9~$T$ (blue circle). Both data sets in the inset can be fit to equation~1 with following parameters: $a_1=0 \Omega$, $a_2=2.57  \Omega/K$, $b_1=3.45\times10^6 \Omega K$ and $b_2=453 K$ (black line) and $a_1= 66.82 \Omega$, $a_2=2.92 \Omega/K$, $b_1=3.45\times10^6 \Omega K$ and $b_2=453 K$ (blue line).}
\label{fig2}
\end{figure}
Whereas the earlier metal-insulator transitions are thought to be universal features in general 2D systems and might all share the same origin associated with interaction and localization physics, the unique features of our observation suggest an entirely different scenario. In particular, the resistance in the `insulating-like' phase is surprisingly low, implying that the apparent negative temperature slope is highly unlikely to be due to localization physics. The quasiparticles are still as mobile as in a metal, in sharp contrast to the many other observations in which the negative temperature slope is regarded as the finite temperature signature of strong localization. Also, the interaction parameter $r_s\sim1.8$ in our 2D system is much smaller than that involved in other 2D MIT experiments, indicating that interaction effects are not germane to our experiments.

In fact, there is strong evidence that our observed ``metal-insulator transition" might be entirely a Fermi surface effect, as predicted in Ref. \cite{Sankar00PRL}. This minimal model based on the Boltzmann transport theory of the Drude model emphasizes the importance of the magneto-orbital coupling without including the spin degree of freedom \cite{Sankar00PRL}. The non-monotonic colossal magnetoresistance in the low temperature limit has already been understood in this framework as the result of a 2D to quasi-3D crossover transition, when the magneto-orbital coupling becomes sufficiently strong \cite{Zhou10PRL}. We note that the magnetic length associated with the applied magnetic field is given by $l_{B}=\frac{26nm}{\sqrt{B}}$, which becomes comparable to the transverse z-width of the quasi-2D wavefunction in our sample around $B=9T$, producing a strong nonperturbative magneto-orbital coupling effect. The temperature dependence of the apparent metal-insulator transition is likely the result of thermal population on a deformed Fermi surface in the presence of a strong parallel field. At low magnetic field where the deformation is not strong and  $T<T_F(B)$, the system behaves as a textbook example of a 2D electron gas with a metallic temperature dependence arising from acoustic phonon scattering. At high magnetic field, the Fermi temperature $T_F(B)$ is suppressed below the temperature $T$.  In this case the electronic system should be viewed more like a low temperature classical plasma rather than a conventional Fermi liquid. Although  a quantitative theory including all possible aspects is currently lacking, the work by Das Sarma and Hwang \cite{Sankar00PRL} predicts, at least qualitatively, a resistivity increasing with decreasing temperature as $R_{xx} \sim T_F/T$ in the classical $T>T_{F}$ high-field regime. In particular, we stress that the fingerprint features of $R_{xx}(B)$ at different temperatures (Fig.~1a) share a close qualitative resemblance with the corresponding theoretical results (Fig.~2 in reference \cite{Sankar00PRL}). In addition, the experimental fact that different $R_{xx}(B)$ curves meet at different temperatures is a distinct and qualitative feature predicted by this theoretical simulation, as the Fermi surface effect is a crossover rather than a sharp transition. This leads us to interpret our observation of the change of sign of $\frac{\partial \rho_{xx}}{\partial T}$ as the result of  the evolution of a quantum metal into a classical plasma caused by the magnetic field induced suppression of $T_F$, leading to the measured temperature dependence exhibiting a competition between the quantum-classical crossover and the acoustic phonon scattering, with the latter always winning at high temperature ($>20K$) independent of whether the system is classical or quantum.

In conclusion, we have observed an apparent metal-insulator transition induced by a parallel magnetic field in a weakly interacting ultraclean 2DEG  with the configuration $\textbf{\emph{I}} \parallel\textbf{\emph{B}}$ . The low-resistivity of our sample with a large metallicity parameter $k_{F}\lambda$,  far from the Ioffe-Regel limit, suggests that this transition is unlikely  to be the result of strong localization. Likewise, the relatively low spin polarization of our system at $\sim$10 T makes it unlikely for spin to play a dominant role. The metallic phase shows a positive temperature dependence fully consistent with the established theory of phonon scattering, while the transition to an `insulating-like' phase is qualitatively described by theoretical simulation of the strong magneto-orbital coupling effect. This phenomenon is not a sharp transition but likely a demonstration of a quantum-classical crossover, where the effective metallic state in the quantum regime gradually evolves into a classical system accompanied by a change of sign of $\frac{\partial \rho_{xx}}{\partial T}$. Our most important discovery in this work is the experimental demonstration that a nominally highly metallic system (with a zero-field low-temperature mean free path approaching a fraction of a mm) could manifest a remarkably strong temperature dependent magnetoresistivity with both positive and negative signs of $\frac{\partial \rho_{xx}}{\partial T}$ (and factors of 2-3 increase or decrease of resistivity as a function of temperature in the 1-10$K$ range) in the presence of an external parallel magnetic field even when the electronic spin degree of freedom plays no role. The inevitable qualitative conclusion following from our work is that the positive or negative sign of the low-temperature resistivity is a poor indicator of an underlying conductor-to-insulator zero-temperature quantum phase transition since the temperature dependence of electronic resistivity is controlled by many mechanisms (e.g. disorder scattering, screening, phonons, magneto-orbital coupling) which may not be controlled by quantum criticality at all.

This work has been supported by NSERC, CIFAR, FQRNT and Microsoft Station-Q. Part of this work has been supported by EuroMagNET under the EU contract no. 228043. The work at Princeton was partially funded by the Gordon and Betty Moore Foundation as well as the National Science Foundation MRSEC Program through the Princeton Center for Complex Materials (DMR-0819860). We also thank W. Knafo, R. Talbot, R. Gagnon and J. Smeros for technical assistance.

\end{document}